\documentclass[aps,prl,twocolumn]{revtex4-2}
\usepackage{hyperref}
\usepackage{graphicx}
\usepackage{physics}
\usepackage{mathrsfs}
\usepackage{amsmath}
\usepackage{amssymb}
\usepackage{amsfonts}
\usepackage{mathtools}
\usepackage{color}
\hypersetup{colorlinks=true,citecolor=blue,linkcolor=blue,urlcolor=blue}

\newcommand{\subheading}[1]{\vspace{10pt}\noindent\textbf{#1}\\[2pt]}
\newcommand{\im}{{\rm i}}
\newcommand{\one}{{\rm I}}
\newcommand{\two}{{\rm I\hspace{-1.2pt}I}}

\begin{document} 
\title{Permutationally Invariant Quantum State Tomography for Fermions} 
\author{Shion Yamashika}
\affiliation{Department of Engineering Science, University of Electro-Communications, Chofu, Tokyo 182-8585, Japan} 
\author{Daisuke Yamamoto}
\affiliation{ Department of Physics, College of Humanities and Sciences, Nihon University, Sakurajosui, Setagaya, Tokyo 156-8550, Japan} 
\affiliation{RIKEN Center for Quantum Computing (RQC), Wako, Saitama 351-0198, Japan}
\affiliation{Global R\&D Center for Business by Quantum-AI Technology (G-QuAT),
National Institute of Advanced Industrial Science and Technology (AIST),
Tsukuba, Ibaraki 305-8568, Japan}

\begin{abstract}
Quantum state tomography provides complete information about a quantum state, but its measurement cost generally grows exponentially with system size. In many-particle quantum simulators, this challenge is further compounded by the limited accessibility of local measurements and controls. Here we develop a tomography protocol for permutation-invariant fermionic many-body states with U(1) particle-number symmetry. We show that any such state is completely determined by the distribution of the total particle number and the occupation of a single collective mode within each particle-number sector, both of which are accessible in current ultracold-atom experiments. The number of required observables scales only linearly with the system size. More generally, the protocol reconstructs the permutation-symmetrized component of arbitrary U(1)-symmetric fermionic states, which can still encode nontrivial many-body and state-level structure beyond conventional few-body observables. We demonstrate this protocol in interacting non-Gaussian states of the complex Sachdev-Ye-Kitaev model and in free-fermion chains across a Lifshitz transition. This framework opens a route toward information-theoretic characterization of strongly correlated itinerant quantum matter in experimentally realistic fermionic quantum simulators. \end{abstract}

\maketitle

\textit{Introduction.---}
Quantum simulators have transformed the study of quantum many-body physics by realizing controllable interacting systems beyond the reach of classical computation or effective theories. This development has proceeded along two complementary directions. One direction is programmable qubit platforms, such as superconducting processors~\cite{Clarke2008,DiCarlo2009,Huang2020,Wu2021,Bravyi2022} and trapped ions~\cite{Johanning2009,Lanyon2011,Blatt2012,Schneider2012,Monroe2013,FossFeig2025}, which offer flexible local control and measurements, making them powerful platforms for controlled many-body dynamics and quantum-information processing. The other is many-particle quantum simulators, especially ultracold atoms in optical lattices~\cite{Bloch2005,Bloch2008b,Lewenstein2012,Zohar2015,Goldman2016,Gross2017,Schfer2020}, which directly realize strongly correlated itinerant quantum matter, a central theme of
condensed-matter physics, under microscopic Hamiltonian control. 

Understanding the many-body phenomena produced in these simulators requires characterizing the underlying quantum states and correlations. In principle, quantum state tomography provides a complete characterization~\cite{Vogel1989,Leonhardt1995,White1999}, since it reconstructs the density matrix and thereby enables the evaluation of arbitrary observables as well as state-level diagnostics such as entanglement entropy. In practice, however, full tomography rapidly becomes infeasible because its measurement cost grows exponentially with system size~\cite{Paris2004,DAriano2003,Dodonov1997,Amiet1999}. 

For qubit systems, substantial progress has been made in mitigating this difficulty. Compressed-sensing tomography and related methods exploit additional assumptions on the state, such as low rank or specific spatial structure~\cite{Gross2010,gross2011,schwemmer2014,kalev2015,Vijayan_2024,liu2011,marmorini2026}. Symmetry-adapted protocols provide another route, as exemplified by permutationally invariant quantum tomography (PIQT)~\cite{Toth2010,Moroder2012,Gao2014}, which reconstructs permutation-invariant many-qubit states from only polynomially many collective measurements. More recently, randomized measurements and classical shadows have emerged as a complementary approach, allowing many observables and state diagnostics to be estimated without reconstructing the full density matrix~\cite{aaronson2018,HuangKuengPreskill2020,elben2022}.

The situation is more restrictive for many-particle quantum simulators. While these platforms have enabled the observation of a wide variety of many-body phenomena~\cite{Bloch2005,Bloch2008b,Lewenstein2012,Zohar2015,Goldman2016,Gross2017,Schfer2020} through measurements of conventional observables such as density profiles, correlation functions, and momentum distributions, access to state-level information remains much more limited. Unlike programmable qubit platforms, they generally do not allow arbitrary local-basis rotations and measurements, preventing the direct implementation of many characterization protocols developed for qubit systems. Recent advances, including measurements of R\'enyi entanglement entropies in bosonic systems~\cite{islam2015,kaufman2016}, demonstrate growing access to quantum-information properties in many-particle systems, yet scalable state characterization remains a major challenge.

In this Letter, we establish a PIQT protocol for itinerant fermionic systems with U(1) particle-number symmetry. Our main result is that the permutation-symmetrized density matrix of any U(1)-symmetric fermionic state is fully reconstructed from only two sets of observables scaling linearly with system size: the total particle-number distribution and the occupation of a single collective mode within each particle-number sector. These quantities are directly accessible through band-mapping measurements~\cite{Bloch2005,Bloch2008b,Lewenstein2012,Zohar2015,Goldman2016,Gross2017,Schfer2020} in ultracold-atom experiments. When the original state is invariant under permutations, the protocol yields a complete reconstruction of the original density matrix. More generally, for states without permutation symmetry, it reconstructs the corresponding permutation-symmetrized density matrix, which can still retain nontrivial state-level structure and correlations beyond conventional observables. After deriving the reconstruction formula, we discuss its utility as a scalable tomography protocol and then demonstrate its broader applicability in interacting non-Gaussian states of the complex Sachdev-Ye-Kitaev (SYK) model and in free-fermion chains across a Lifshitz transition.

\textit{Preliminaries.---}
We consider a system of $N$ fermionic modes. Here, a mode denotes a single-particle mode in a chosen basis; it may correspond to a lattice site, an internal state, or another mode label. We denote the annihilation and creation operators in the $i$-th mode by $c_i$ and $c_i^\dag$, respectively. Throughout this work, we focus on a state $\rho$ with the U(1) symmetry generated by the total particle number operator $Q=\sum_{i=1}^N c_i^\dag c_i$. This assumption is natural for fermionic quantum simulators with cold atoms, where the total particle number is conserved to an excellent approximation and measurements are usually performed in a fixed or resolved particle-number sector~\cite{Bloch2005,Bloch2008b,Lewenstein2012,Zohar2015,Goldman2016,Gross2017,Schfer2020}. The U(1) symmetry forbids coherences between different particle-number sectors and hence the density matrix can be written as 
\begin{align}\label{eq:rho block}
\rho = \bigoplus_{q=0}^N p_q \rho_q. 
\end{align}
Here, $p_q =\Tr_{}[P_q\rho]$ is the probability of finding $q$ particles, $P_q$ is the projector onto the $q$-particle sector, and $\rho_q = P_q \rho P_q/p_q$ is the normalized density matrix in that sector. 

Such a fermionic state is specified by the expectation values of many-body correlation functions such as $c_i^\dag c_j^\dag c_k c_l$, and their higher-order generalizations, whose number grows exponentially with system size. Full reconstruction of the density matrix would therefore require measuring a complete set of high-order fermionic correlations, which is generally unavailable in many-particle platforms such as ultracold atoms.

Rather than attempting such a full reconstruction, we target the permutation-symmetrized version of the original density matrix defined as follows. Let $\mathcal{S}_N$ denote the permutation group of $N$ modes. For a permutation $\pi\in \mathcal{S}_N$, we define the corresponding unitary operator $U_\pi$ by 
\begin{align}
U_\pi c_i U_\pi^\dag =c_{\pi(i)}. 
\end{align}
The permutation-symmetrized density matrix is then defined as 
\begin{align}\label{eq:Pi(rho)}
\Pi(\rho)=\frac{1}{N!} \sum_{\pi \in \mathcal{S}_N} U_\pi \rho U_\pi^\dag. 
\end{align}
This operation preserves positivity, $\Pi(\rho)\geq 0$, and trace $\Tr \Pi(\rho)=1$, and hence $\Pi(\rho)$ is again a density matrix. By construction, $\Pi(\rho)$ is invariant under all permutations, i.e., $U_\pi \Pi(\rho) U_\pi^\dag=\Pi (\rho)$ for every $\pi \in \mathcal{S}_N$. 

\begin{figure*}
\centering
\includegraphics[width=0.95\textwidth]{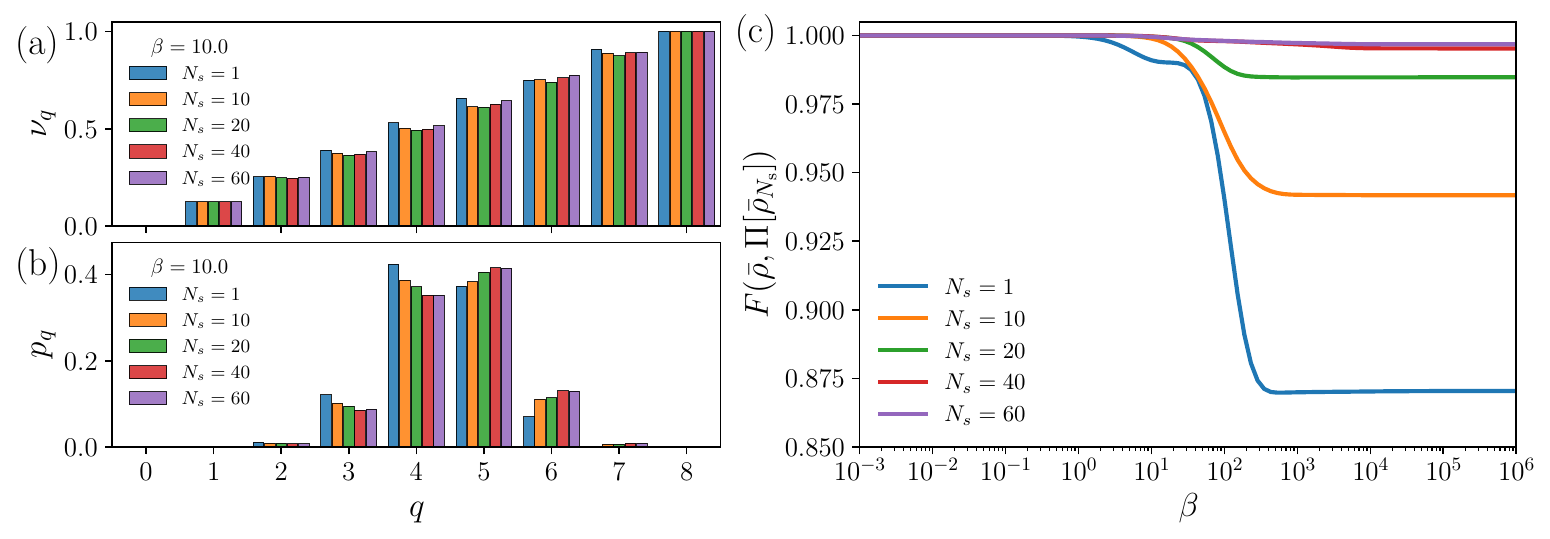}
\caption{Few-sample construction of a permutation-invariant disorder-averaged state in the complex SYK model. 
(a,b) Conditional uniform-mode occupation $\nu_q$ and particle-number distribution $p_q$ at $\beta=10$, estimated from $N_{\rm s}$ disorder realizations. These two quantities are the only inputs inserted into Eq.~\eqref{eq:PIQT} to construct $\Pi(\bar{\rho}_{N_{\rm s}}(\beta))$. 
(c) Uhlmann fidelity between $\Pi(\bar{\rho}_{N_{\rm s}}(\beta))$ and a large-sample reference for $\bar{\rho}(\beta)$ obtained from $10^4$ disorder realizations. 
We set $N=8$ in all panels.}\label{fig:cSYK}
\end{figure*}

\textit{Fermionic PIQT.---}
The central result of this work is that the permutation-symmetrized density matrix in Eq.~\eqref{eq:Pi(rho)} can be written as 
\begin{align}\label{eq:PIQT}
\Pi(\rho) = \bigoplus_{q=0}^{N} p_q 
\qty(
\frac{1-\nu_q}{\binom{N-1}{q}} P_{q,0}
+ \frac{\nu_q}{\binom{N-1}{q-1}} P_{q,1}). 
\end{align}
Here, $\nu_q=\Tr_{}[\rho_q n_0]$ is the conditional occupation of the uniform mode in the $q$-particle sector, where $n_0=d_0^\dag d_0$ and $d_k=N^{-1/2}\sum_{i=1}^N e^{-\im k i} c_i$. $P_{q,0}=P_q(1-n_0)$ and $P_{q,1}=P_q n_0$ are the projection operators onto the subspaces of the $q$-particle sector in which the uniform mode is empty and occupied, respectively. Terms associated with zero-dimensional subspaces are omitted: only $P_{0,0}$ contributes for $q=0$, and only $P_{N,1}$ contributes for $q=N$. Equation~\eqref{eq:PIQT} is not restricted to Gaussian states; it holds for any density matrix with U(1) particle-number symmetry, including interacting non-Gaussian states.

\textit{Proof of Eq.~\eqref{eq:PIQT}.---}
Because the permutations conserve the total particle number, permutation symmetrization $\Pi(\cdot)$ acts independently in each $q$-particle sector of the density matrix~\eqref{eq:rho block}. Let $\mathcal H_q$ be the $q$-particle Hilbert space and let $\Pi_q$ denote permutation symmetrization restricted to $\mathcal H_q$. Equation~\eqref{eq:rho block} then gives
\begin{align}\label{eq:Pi_sector}
\Pi(\rho)=\bigoplus_{q=0}^N p_q\Pi_q(\rho_q).
\end{align}
The proof of Eq.~\eqref{eq:PIQT} thus reduces to determining $\Pi_q(\rho_q)$ for a fixed $q$. The sectors $q=0$ and $q=N$ are one-dimensional and satisfy Eq.~\eqref{eq:PIQT} immediately, and we therefore restrict the following argument to $1\leq q\leq N-1$.

Permutations conserve not only the total particle number but also the uniform-mode occupation $n_0$. This fact allows us to decompose the representation of $\mathcal S_N$ on $\mathcal H_q$ into the two subspaces $V_{q,n}=P_{q,n}\mathcal H_q$ with $n=0,1$. 
More specifically, if we rewrite the one-particle Hilbert space as
\begin{align}
\mathcal H_1
=
\mathrm{span}\{d_0^\dag\ket{0}\}
\oplus W,
\end{align}
where $W=\mathrm{span}\{d_k^\dag \ket{0}\,|k\neq0\}$ is the subspace orthogonal to the uniform mode, we obtain 
\begin{align}\label{eq:H_q}
\mathcal H_q
=V_{q,0}\oplus V_{q,1}, 
\end{align}
with 
\begin{gather}
V_{q,0}=\wedge^q W,~
V_{q,1}=
\mathrm{span}\{d_0^\dag\ket{0}\}\wedge \wedge^{q-1}W.
\end{gather}
Here, $\wedge^q$ denotes the $q$th exterior power. Since every permutation acts trivially on $\mathrm{span}\{d_0^\dag \ket{0}\}$, $V_{q,n}$ in Eq.~\eqref{eq:H_q} carries the representation $\wedge^{q-n} W$ ($n=0,1$). These two representations are irreducible~\cite{Fulton2004} and mutually inequivalent, as shown in the End Matter.

Applying the decomposition~\eqref{eq:H_q} to the density matrix $\rho_q$ and taking the permutation symmetrization, we obtain 
\begin{align}\label{eq:Pi(rho_q)}
\Pi_q(\rho_q)=\sum_{n,m=0}^1X_{q,nm},
\end{align}
where $X_{q,nm}=\Pi_q\qty(P_{q,n}\rho_qP_{q,m})$.
By definition, $X_{q,nm}$ is invariant under every permutation, implying that 
\begin{align}\label{eq:intertwiner}
U_{\pi,q,n}X_{q,nm}
=X_{q,nm}U_{\pi,q,m}
\qquad \forall \pi\in\mathcal S_N,
\end{align}
where $U_{\pi,q,n}$ is the restriction of $U_\pi$ to the subspace $V_{q,n}$. 
For $n\neq m$, $X_{q,nm}$ in Eq.~\eqref{eq:intertwiner} is an intertwiner between inequivalent irreducible representations $\wedge^{q}W$ and $\wedge^{q-1}W$. Schur's lemma then gives $X_{q,nm}=0$. For $n=m$, $X_{q,nn}$ commutes with an irreducible representation, and hence Schur's lemma gives $X_{q,nn}=\alpha_{q,n}P_{q,n}$, where $\alpha_{q,n}$ are constants. Equation~\eqref{eq:Pi(rho_q)} therefore reduces to 
\begin{align}
\Pi_q(\rho_q)
=
\alpha_{q,0}P_{q,0}
+
\alpha_{q,1}P_{q,1}.
\end{align}
The specific forms of the coefficients $\alpha_{q,n}$ can be derived from the conditions $\Tr[\Pi_q(\rho_q)]=1$ and $\Tr[\Pi_q(\rho_q)n_0]=\nu_q$ as 
\begin{align}
\alpha_{q,0}=\frac{1-\nu_q}{\mqty(N-1 \\q)},
\qquad
\alpha_{q,1}=\frac{\nu_q}{\mqty(N-1 \\ q-1)}.
\end{align}
Multiplying by the sector probability $p_q$ and summing over $q$ gives Eq.~\eqref{eq:PIQT}.

\textit{Applications.---}
Equation~\eqref{eq:PIQT} has two immediate uses. First, if the original state is permutationally invariant, it gives a scalable tomography protocol: For $\rho=\Pi(\rho)$, the density matrix can be reconstructed from the input data $\{p_q,\nu_q\}_{q=0}^{N}$, whose number grows only linearly with $N$. 
Second, Eq.~\eqref{eq:PIQT} gives expectation values of permutation-invariant observables even when the original state is not permutation invariant: Whenever $\rho$ respects the U(1) particle-number symmetry, $\Pi(\rho)$ can always be reconstructed from Eq.~\eqref{eq:PIQT}. For any observable $O$ satisfying $\Pi(O)=O$, one then obtains $\Tr_{}[\rho O]=\Tr[\rho \Pi(O)]=\Tr[\Pi(\rho)O]$. 

We now illustrate the broader utility of Eq.~\eqref{eq:PIQT} in two physically distinct settings.

\textit{Example 1: Complex SYK model.---}
We first consider the complex SYK model~\cite{Sachdev2015,Gu2020}
\begin{align}
H_J = \sum_{i<j,k<l} J_{ij,kl}c_i^\dag c_j^\dag c_k c_l, 
\end{align}
where the couplings $J_{ij,kl}$ are complex Gaussian variables with zero mean and variance $N^{-3}$ and satisfy $J_{ij,kl}=-J_{ji,kl}=-J_{ij,lk}=J_{kl,ij}^*$.
Cold-atom realizations of this model and closely related variants have been proposed~\cite{danshita2017,wei2021,creffield2026}.
We aim to reconstruct the disorder-averaged Gibbs state
\begin{align}\label{eq:bar rho}
\bar{\rho}(\beta) = \mathbb{E}_J[\rho_J(\beta)],
\end{align}
where $\rho_J(\beta)=e^{-\beta H_J}/\Tr_{}(e^{-\beta H_J})$ is the Gibbs state for a realization $J$ at the inverse temperature $\beta$. 
Since the coupling distribution is invariant under permutations of the mode labels, $\bar{\rho}(\beta)$ is permutation invariant: $\Pi(\bar{\rho}(\beta))=\bar{\rho}(\beta)$.
The difficulty in constructing $\bar{\rho}(\beta)$ is that it requires averaging over infinitely many disorder realizations, which is impractical in both numerical calculations and experiments.

Figure~\ref{fig:cSYK} shows that Eq.~\eqref{eq:PIQT} allows us to circumvent this difficulty. Here, we estimate $\nu_q$ and $p_q$ from $N_{\rm s}$ realizations, as shown in Figs.~\ref{fig:cSYK}(a) and \ref{fig:cSYK}(b). We then insert them into Eq.~\eqref{eq:PIQT} to construct $\Pi(\bar{\rho}_{N_{\rm s}}(\beta))$, where $\bar{\rho}_{N_{\rm s}}(\beta)=N_{\rm s}^{-1}\sum_{a=1}^{N_{\rm s}}\rho_{J_a}(\beta)$ is the finite-sample averaged state. Figure~\ref{fig:cSYK}(c) plots the Uhlmann fidelity $F(\rho,\sigma)=\Tr\sqrt{\sqrt{\rho}\sigma\sqrt{\rho}}$~\cite{Uhlmann2010,Jozsa1994} between $\Pi(\bar{\rho}_{N_{\rm s}}(\beta))$ and $\bar{\rho}(\beta)$ as a function of $\beta$. The fidelity remains close to unity even with only a few tens of realizations, showing that the finite-sample construction approximates well $\bar{\rho}(\beta)$.

The effectiveness of this construction comes from the additional averaging introduced by permutation symmetrization, which gives
\begin{align}\label{eq:orbit average}
\Pi(\bar{\rho}_{N_{\rm s}}(\beta))
=\frac{1}{N_{\rm s}N!}\sum_{a=1}^{N_{\rm s}}\sum_{\pi\in\mathcal S_N}
U_\pi\rho_{J_a}(\beta)U_\pi^\dag.
\end{align}
Each state $U_\pi\rho_{J_a}(\beta)U_\pi^\dag$ in the above equation is the Gibbs state obtained by applying the same permutation to the coupling labels. Equation~\eqref{eq:orbit average} therefore shows that the reconstructed state incorporates the entire permutation orbit of every sampled realization without explicitly sampling additional disorder realizations, effectively increasing the number of samples from $N_{\rm s}$ to $N_{\rm s}N!$. In addition, the reconstructed state respects the permutation symmetry of $\bar{\rho}(\beta)$, thereby removing the permutation-noninvariant component of the finite-sample error. These facts explain the high fidelity obtained from only a few realizations.

\textit{Example 2: Lifshitz transition.---}
We next consider a Lifshitz transition in a one-dimensional free-fermion ground state corresponding to the Fermi sea 
\begin{align}
\ket{\Psi}=\prod_{k:\varepsilon_k<0}d_k^\dag \ket{0}.
\end{align}
Here $\varepsilon_k$ is the single-particle dispersion. A Lifshitz transition occurs when varying $\varepsilon_k$ changes the topology of the occupied momentum region~\cite{Lifshitz1960,Blanter1994}.
It is known that, for an interval $A$ of $N_A$ sites, the ground-state entanglement entropy $S(\rho_A)=-\Tr_A \rho_A\log \rho_A$ of the reduced state $\rho_A=\Tr_{\bar A}\ketbra{\Psi}$ exhibits a nonanalyticity at a Lifshitz transition~\cite{Rodney2013,Peschel2009}. We ask whether this signature survives permutation symmetrization. 

Since $\rho_A$ has U(1) symmetry, $\Pi_A(\rho_A)$ can be obtained by Eq.~\eqref{eq:PIQT}, where $\Pi_A$ stands for the permutation symmetrization with respect to the sites inside subsystem $A$. As derived in the Supplemental Material (SM)~\cite{SM}, its von Neumann entropy can be calculated as 
\begin{align}
S(\Pi_A(\rho_A))= N_A h(\varrho) -\frac{1}{2}\ln N_A  +O(\ln \ln N_A),
\label{eq:S_Lifshitz}
\end{align}
where $\varrho=\sum_k\Theta(-\varepsilon_k)/N$ and $h(\varrho)=-\varrho\ln\varrho-(1-\varrho)\ln(1-\varrho)$. At a Lifshitz transition point, a Fermi pocket appears or disappears, causing the fraction $\varrho$ of occupied modes to change nonanalytically. The function $h(\varrho)$ inherits this nonanalyticity. Equation~\eqref{eq:S_Lifshitz} therefore predicts a nonanalyticity in $S(\Pi_A(\rho_A))$ at the transition point.

\begin{figure}
\raggedright
\includegraphics[width=0.48\textwidth]{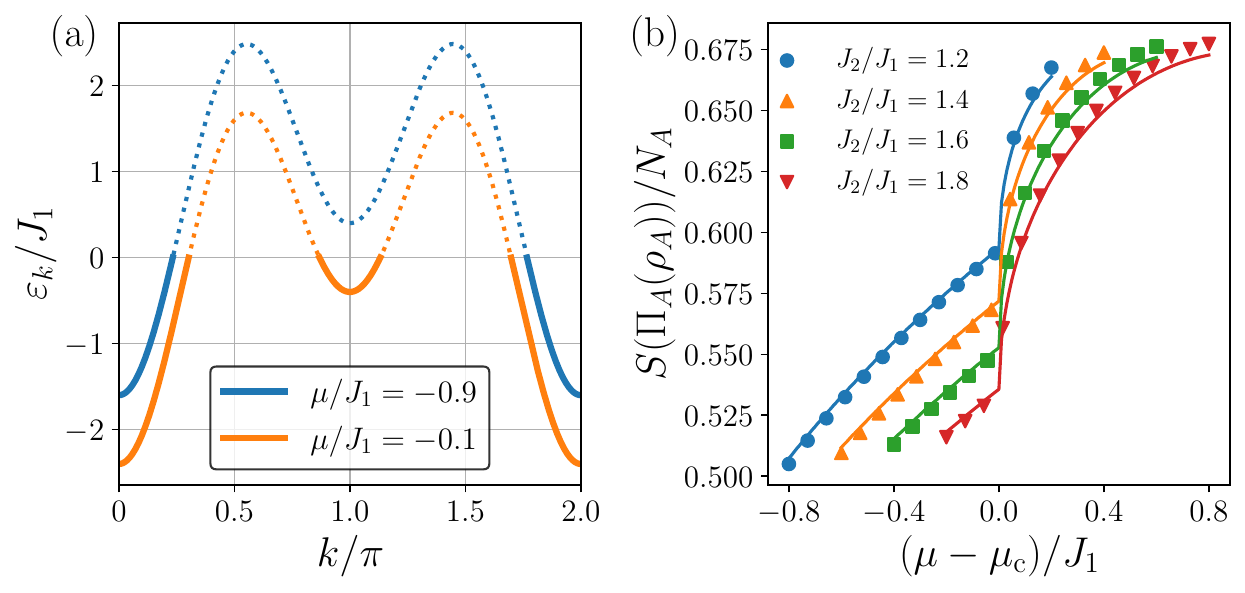}
\caption{(a) The energy dispersion relation of the Hamiltonian~\eqref{eq:H_J1J2} with $J_2/J_1=1.5$. The empty and occupied modes are plotted by dotted and solid lines, respectively. 
(b) von Neumann entropy of the permutation-symmetrized reduced density matrix for the ground state of the Hamiltonian~\eqref{eq:H_J1J2}. The symbols and solid lines are the exact numerical results and the asymptotic prediction in Eq.~\eqref{eq:S_Lifshitz}, respectively. We use $N_A=2^7$ and $N=2^{14}$.} \label{fig:Lifshitz}
\end{figure}

We now illustrate this prediction with the nearest- and next-nearest-neighbor hopping model
\begin{align}\label{eq:H_J1J2}
H=-\frac{1}{2}\sum_i[J_1c_i^\dag c_{i+1}+J_2c_i^\dag c_{i+2}+\mathrm{H.c.}]-\mu \sum_i c_i^\dag c_i, 
\end{align}
where the single-particle dispersion relation is given by
\begin{align}
\varepsilon_k = -J_1\cos k -J_2 \cos 2k-\mu. 
\end{align}
For $J_2>J_1/4$, the local minimum at $k=\pi$ crosses the Fermi level at $\mu_{\rm c}=J_1-J_2$, creating a Fermi pocket for $\mu>\mu_{\rm c}$ [see Fig.~\ref{fig:Lifshitz}(a)]. As shown in Fig.~\ref{fig:Lifshitz}(b), $S(\Pi_A(\rho_A))$ indeed exhibits the predicted nonanalyticity at the transition point.

\textit{Experimental Implementation.---}
As mentioned before, the fermionic PIQT protocol based on Eq.~\eqref{eq:PIQT} requires only two types of data in each experimental shot: the total particle number and the occupation of the uniform mode. For fermions in an optical lattice, the latter corresponds to the zero-momentum mode. Repeated band-mapping measurements therefore give both quantities simultaneously~\cite{Bloch2005,Bloch2008b,Lewenstein2012,Zohar2015,Goldman2016,Gross2017,Schfer2020}: the histogram of the measured total particle number gives $p_q$, while the conditional average of the zero-momentum occupation among shots with total particle number $q$ gives $\nu_q$. Once these quantities are obtained for all $q$, the permutation-symmetrized density matrix is reconstructed directly from Eq.~\eqref{eq:PIQT}.
Thus, the fermionic PIQT protocol does not require local basis rotations, site-by-site control, or measurements of a complete set of high-order fermionic correlations. It relies only on number- and momentum-resolved data, making it well matched to ultracold-atom fermionic simulators.

\textit{Conclusions.---}
We have shown that the permutation-symmetrized density matrix of a U(1)-symmetric fermionic many-body state is generally determined by particle-number statistics and one uniform-mode occupation in each particle-number sector. This gives a fermionic version of PIQT protocol in which the quantities to be measured scale linearly with system size and are accessible in ultracold-atom experiments. The formula applies beyond Gaussian states, as illustrated by the disorder-averaged Gibbs state of the complex SYK model, and it can also extract transition signatures from states that are not permutation invariant, as shown for spatial subsystems of free-fermion chains across Lifshitz transitions. These results establish fermionic PIQT as an efficient route to information-theoretic characterization of many-body fermionic states without full density-matrix reconstruction.

Looking ahead, it will be important to extend the present framework beyond U(1)-symmetric fermionic states and to bosonic systems. Such extensions would broaden the range of quantum simulators to which PIQT can be applied. Another important direction is to clarify what kinds of quantum information are retained in the permutation-symmetrized density matrix. Understanding how this reconstructed state reflects strong correlations, topology, and other organizing principles of quantum matter would help turn the present protocol into a powerful diagnostic for many-particle quantum simulators.

~

\textit{Acknowledgments---} 
We thank T. Fukuhara and H. Katsura for fruitful discussions. 
This work was supported by JSPS KAKENHI Grant Numbers JP25K23355 (SY), JP26K17050 (SY), JP23K25830 (DY), JP24K06890 (DY), and JP26K00664 (DY), and JST PRESTO Grant Number~JPMJPR245D (DY). SY acknowledges support from the University of Electro-Communications.

\bibliographystyle{./apsrev4-2-title}
\bibliography{ref} 

~

\begin{center}
\textbf{\Large End Matter}
\end{center}
Here we verify the claim that the two nonzero irreducible representations carried by $V_{q,0}$ and $V_{q,1}$ are inequivalent for $1\leq q\leq N-1$. As established in Eq.~\eqref{eq:H_q}, these representations are $\wedge^q W$ and $\wedge^{q-1}W$, respectively. For $N\neq 2q$, they are trivially inequivalent because their dimensions 
\begin{align}
    \dim \wedge^{q-n} W=\mqty(N-1 \\q-n), \quad (n=0,1)
\end{align} 
are different.

It remains to check the half-filled case $N=2q$, where $\dim \wedge^{q}W=\dim \wedge^{q-1}W$. 
To this end, let us consider the transposition $\tau\in \mathcal{S}_N$ such that $\tau(1)=2$, $\tau(2)=1$, and $\tau(i)=i$ for $i=3,4,\ldots,N$. 
In the standard representation $W$, this transposition has eigenvalue $+1$ with multiplicity $N-2$ and eigenvalue $-1$ with multiplicity $1$. 
Therefore, on $\wedge^r W$, the character of $\tau$ reads 
\begin{align}
    \Tr_{\wedge^r W}(\tau)
    =
    \binom{N-2}{r}
    -
    \binom{N-2}{r-1}.
\end{align}
Indeed, the first term counts basis vectors of $\wedge^r W$ that do not contain the $-1$ eigenvector of $\tau$, while the second term counts those that contain it. 
If two representations are equivalent, their characters must agree for every group element. 
At $N=2q$, however,
\begin{gather}
    \Tr_{\wedge^q W}(\tau)
    =
    \binom{2q-2}{q}
    -
    \binom{2q-2}{q-1}, \\
    \Tr_{\wedge^{q-1} W}(\tau)
    =
    \binom{2q-2}{q-1}
    -
    \binom{2q-2}{q-2},
\end{gather}
and these two numbers are different. 
Thus $\wedge^q W$ and $\wedge^{q-1}W$ are inequivalent even when $N=2q$. 

\clearpage
\def\includedinsupplement{}
\ifdefined\includedinsupplement
\else
\documentclass[onecolumn,superscriptaddress]{revtex4-2}
\usepackage{xr}
\externaldocument[main-][nocite]{main}
\usepackage{hyperref}
\usepackage{graphicx}
\usepackage{physics}
\usepackage{mathrsfs}
\usepackage{amsmath}
\usepackage{amssymb}
\usepackage{amsfonts}
\usepackage{mathtools}
\usepackage{color}
\hypersetup{colorlinks=true,citecolor=blue,linkcolor=blue,urlcolor=blue}

\newcommand{\subheading}[1]{\vspace{10pt}\noindent\textbf{#1}\\[2pt]}
\newcommand{\im}{{\rm i}}
\newcommand{\one}{{\rm I}}
\newcommand{\two}{{\rm I\hspace{-1.2pt}I}}

\begin{document} 
\fi

\ifdefined\includedinsupplement
\newcommand{\mainEqref}[1]{\eqref{#1}}
\else
\newcommand{\mainEqref}[1]{\eqref{main-#1}}
\fi

\ifdefined\includedinsupplement
\else
\setcounter{page}{1}
\fi
\onecolumngrid
\newcounter{equationSM}
\newcounter{figureSM}
\newcounter{tableSM}
\stepcounter{equationSM}
\setcounter{equation}{0}
\setcounter{figure}{0}
\setcounter{table}{0}
\setcounter{section}{0}
\renewcommand{\theequation}{\textsc{sm}-\arabic{equation}}
\renewcommand{\thefigure}{\textsc{sm}-\arabic{figure}}
\renewcommand{\thetable}{\textsc{sm}-\arabic{table}}
\renewcommand{\theHequation}{SM.\arabic{equation}}
\renewcommand{\theHfigure}{SM.\arabic{figure}}
\renewcommand{\theHtable}{SM.\arabic{table}}

\begin{center}
  {\Large{\bf Supplemental Material}}
\end{center} 
\tableofcontents

\section{Von Neumann entropy of the permutation-symmetrized reduced density matrix for a Fermi sea}
Here, we give the derivation of Eq.~\mainEqref{eq:S_Lifshitz} in the main text. 
We consider a one-dimensional free-fermion ground state
\begin{align}
    \ket{\Psi}
    =
    \prod_{k:\varepsilon_k<0}
    d_k^\dagger
    \ket{\Omega}.
\end{align}
We divide the whole system into subsystem $A$ consisting of $N_A$ consecutive sites and the rest. The reduced density matrix for subsystem $A$ is
\begin{align}
    \rho_A
    =
    \Tr_{\bar A}    \ketbra{\Psi}. 
\end{align}
It has U(1) symmetry with respect to the subsystem particle-number operator
\begin{align}
    Q_A
    =
    \sum_{i\in A}
    c_i^\dagger c_i .
\end{align}
We denote by \(p_q\) the probability of finding \(q\) particles in \(A\),
\begin{align}
    p_q
    =
    \Tr(P_q\rho_A),
\end{align}
where \(P_q\) is the projector onto the \(q\)-particle sector of the subsystem.

According to Eq.~\mainEqref{eq:PIQT}, the permutation-symmetrized reduced density matrix can be written as 
\begin{align}
    \Pi_A(\rho_A)
    =
    \bigoplus_{q=0}^{N_A}
    p_q
    \left[
    \frac{1-\nu_q}{D_{q,0}} P_{q,0}
    +
    \frac{\nu_q}{D_{q,1}} P_{q,1}
    \right],
    \label{eq:SM_PIQT_subsystem}
\end{align}
where
\begin{align}
    D_{q,0}
    =
    \binom{N_A-1}{q},
    \qquad
    D_{q,1}
    =
    \binom{N_A-1}{q-1}.
\end{align}
Here \(P_{q,0}\) and \(P_{q,1}\) project onto the subspaces with \(q\) particles and with the uniform orbital empty or occupied, respectively. The number
\begin{align}
    \nu_q
    =
    \Tr(\rho_{A,q} n_{A,0})
\end{align}
is the occupation of the uniform orbital conditioned on the \(q\)-particle sector, with
\begin{align}
    n_{A,0}
    =
    d_{A,0}^\dagger d_{A,0},
    \qquad
    d_{A,0}
    =
    \frac{1}{\sqrt{N_A}}
    \sum_{i\in A}
    c_i .
\end{align}
The eigenvalues of \(\Pi_A(\rho_A)\) are therefore
\begin{align}
    \lambda_{q,0}
    =
    \frac{p_q(1-\nu_q)}{D_{q,0}},
    \qquad
    \lambda_{q,1}
    =
    \frac{p_q\nu_q}{D_{q,1}},
\end{align}
with degeneracies \(D_{q,0}\) and \(D_{q,1}\), respectively. Hence, the von Neumann entropy of $\Pi_A(\rho_A)$ can be written as 
\begin{align}
    S(\Pi_A(\rho_A))
    =
    -\sum_{q=0}^{N_A}
    p_q \ln p_q
    +
    \sum_{q=0}^{N_A}
    p_q
    \left[
    (1-\nu_q)\ln D_{q,0}
    +
    \nu_q \ln D_{q,1}
    \right]
    +
    \sum_{q=0}^{N_A}
    p_q h(\nu_q),
    \label{eq:SM_entropy_exact}
\end{align}
where
\begin{align}
    h(\nu)
    =
    -\nu\ln\nu
    -
    (1-\nu)\ln(1-\nu).
\end{align}
The last term in Eq.~\eqref{eq:SM_entropy_exact} is bounded by \(\ln 2\), and therefore contributes only \(O(1)\) to the entropy.

Using
\begin{align}
    D_{q,0}
    =
    \left(1-\frac{q}{N_A}\right)
    \binom{N_A}{q},
    \qquad
    D_{q,1}
    =
    \frac{q}{N_A}
    \binom{N_A}{q},
\end{align}
we can rewrite the second term in Eq.~\eqref{eq:SM_entropy_exact} as
\begin{align}
    \sum_q p_q \ln\binom{N_A}{q}
    +
    \sum_q p_q
    \left[
    (1-\nu_q)\ln\left(1-\frac{q}{N_A}\right)
    +
    \nu_q\ln\left(\frac{q}{N_A}\right)
    \right].
\end{align}
The second sum is at most \(O(1)\). Therefore, up to \(O(1)\) terms, we obtain 
\begin{align}
    S(\Pi_A(\rho_A))
    =
    -\sum_q p_q\ln p_q
    +
    \sum_q p_q
    \ln\binom{N_A}{q}
    +
    O(1).
    \label{eq:SM_entropy_reduced}
\end{align}

We now evaluate the two terms in Eq.~\eqref{eq:SM_entropy_reduced}. The particle-number distribution \(p_q\) is obtained from the full counting statistics of \(Q_A\)~\cite{Levitov1993}:
\begin{align}
    p_q
    =
    \int_{-\pi}^{\pi}
    \frac{d\theta}{2\pi}
    e^{-iq\theta}
    \chi_A(\theta),
    \qquad
    \chi_A(\theta)
    =
    \Tr\left[
    \rho_A e^{i\theta Q_A}
    \right].
\end{align}
For a Gaussian fermionic state, the full-counting statistics can be written as~\cite{Klich2002}
\begin{align}
    \chi_A(\theta)
    =
    \det
    \left[
    1-C+e^{i\theta}C
    \right],
    \label{eq:SM_FCS_det}
\end{align}
where
\begin{align}\label{eq:C_ij}
    C_{ij}
    =
    \Tr(\rho_A c_i^\dagger c_j),
    \qquad i,j\in A,
\end{align}
is the correlation matrix of the subsystem.

For a translation-invariant Fermi sea, the two-point correlation matrix reads 
\begin{align}\label{eq:C_ij_Toepitz}
    C_{ij}
    =
    \int_{-\pi}^{\pi}
    \frac{dk}{2\pi}
    e^{ik(j-i)}
    \Theta[-\varepsilon_k].
\end{align}
Thus, the determinant in Eq.~\eqref{eq:SM_FCS_det} is a Toeplitz determinant with symbol
\begin{align}
    f_\theta(k)
    =
    1+(e^{i\theta}-1)\Theta[-\varepsilon_k].
\end{align}
The symbol becomes nonanalytic at the Fermi points, at which $\varepsilon_k=0$. If $N_F$ denotes the number of Fermi points, the Fisher-Hartwig theorem gives~\cite{Fisher1969,Ivanov2013}, for fixed $\theta\in(-\pi,\pi)$,
\begin{align}
    \ln \chi_A(\theta)
    =
    i\theta N_A\varrho
    -
    \frac{N_F\theta^2}{4\pi^2}
    \ln N_A
    +
    O(1),
    \label{eq:SM_FCS_asymptotic}
\end{align}
where
\begin{align}
    \varrho
    =
    \int_{-\pi}^{\pi}
    \frac{dk}{2\pi}
    \Theta[-\varepsilon_k]
\end{align}
is the filling of the Fermi sea.

Equation~\eqref{eq:SM_FCS_asymptotic} shows that \(p_q\) is asymptotically Gaussian:
\begin{align}
    p_q
    =
    \frac{1}{\sqrt{2\pi V_A}}
    \exp
    \left[
    -\frac{(q-N_A\varrho)^2}{2V_A}
    \right]
    +o(1),
    \qquad
    V_A
    =
    \frac{N_F}{2\pi^2}
    \ln N_A
    +
    O(1).
    \label{eq:SM_pq_gaussian}
\end{align}
The entropy of this Gaussian distribution is
\begin{align}
    -\sum_q p_q\ln p_q
    =
    \frac{1}{2}\ln(2\pi e V_A)
    +
    O(1)
    =
    \frac{1}{2}\ln\ln N_A
    +
    O(1).
    \label{eq:SM_particle_entropy}
\end{align}

Next, we evaluate the binomial term in Eq.~\eqref{eq:SM_entropy_reduced}. Since \(p_q\) is concentrated around \(q=N_A\varrho\) with variance \(V_A=O(\ln N_A)\), we use Stirling's formula at \(q/N_A=\varrho+O(\sqrt{\ln N_A}/N_A)\):
\begin{align}
    \ln\binom{N_A}{q}
    =
    N_A h\left(\frac{q}{N_A}\right)
    -
    \frac{1}{2}
    \ln
    \left[
    2\pi N_A
    \frac{q}{N_A}
    \left(1-\frac{q}{N_A}\right)
    \right]
    +
    O(N_A^{-1}).
\end{align}
Averaging it with the weight \(p_q\), we obtain 
\begin{align}
    \sum_q p_q \ln\binom{N_A}{q}
    =
    N_A h(\varrho)
    -
    \frac{1}{2}\ln N_A
    +
    O(1).
    \label{eq:SM_binomial_average}
\end{align}
The correction from expanding \(h(q/N_A)\) around \(q/N_A=\varrho\) is \(O(V_A/N_A)\), and is therefore included in the \(O(1)\) term.

Combining Eqs.~\eqref{eq:SM_particle_entropy} and \eqref{eq:SM_binomial_average}, we obtain Eq.~\mainEqref{eq:S_Lifshitz} as 
\begin{align}
    S(\Pi_A(\rho_A))
    =
    N_A h(\varrho)
    -
    \frac{1}{2}\ln N_A
    +
    \frac{1}{2}\ln\ln N_A
    +
    O(1).
    \label{eq:SM_entropy_final_precise}
\end{align}

\section{Permutation-symmetrized density matrix of Gaussian states}

Here, we describe the method used to numerically calculate the von Neumann entropy in Eq.~\eqref{eq:SM_entropy_exact} from the two-point correlation matrix given by Eq.~\eqref{eq:C_ij}. 

To calculate Eq.~\eqref{eq:SM_entropy_exact}, we need $\{p_q,\nu_q\}_{q=0}^{N_A}$. To this end, we first diagonalize this matrix as
\begin{align}
    C_{ij}
    =
    \sum_{\alpha=1}^{N_A}
    u_{\alpha i}
    \lambda_\alpha
    u_{\alpha j}^{*},
\end{align}
where \(0\le \lambda_\alpha\le 1\). We define fermionic normal modes
\begin{align}
    f_\alpha
    =
    \sum_{i\in A}
    u_{\alpha i} c_i .
\end{align}
In this basis, the reduced density matrix can be written as~\cite{Peschel2003}
\begin{align}
    \rho_A
    =
    \sum_{\mathbf n}
    p_{\mathbf n}
    \ketbra{\mathbf n}
    \qquad
    p_{\mathbf n}
    =
    \prod_{\alpha=1}^{N_A}
    \lambda_\alpha^{n_\alpha}
    (1-\lambda_\alpha)^{1-n_\alpha},
    \label{eq:SM_rhoA_diag}
\end{align}
where \(\mathbf n=(n_1,\ldots,n_{N_A})\), \(n_\alpha\in\{0,1\}\), and
\begin{align}
    \ket{\mathbf n}
    =
    \prod_{\alpha=1}^{N_A}
    (f_\alpha^\dagger)^{n_\alpha}
    \ket{\Omega}.
\end{align}

The probability \(p_q\) of finding \(q\) particles in \(A\) is
\begin{align}
    p_q
    =
    \sum_{\mathbf{n}:|\mathbf{n}|=q}
    p_{\mathbf n}.
\end{align}
We can compute these probabilities from the generating polynomial
\begin{align}
    F(z)
    =
    \prod_{\alpha=1}^{N_A}
    \left[
    (1-\lambda_\alpha)+\lambda_\alpha z
    \right]
    =
    \sum_{q=0}^{N_A}
    p_q z^q.
    \label{eq:SM_generating_pq}
\end{align}
Equivalently, we use the recurrence
\begin{align}
    p_q^{(r)}
    =
    (1-\lambda_r)p_q^{(r-1)}
    +
    \lambda_r p_{q-1}^{(r-1)},
    \label{eq:SM_pq_recursion}
\end{align}
Here, \(p_q^{(r)}\) is the probability of finding \(q\) particles among the first \(r\) eigenmodes. We obtain the probability \(p_q=p_q^{(N_A)}\) by numerically solving the recursion relation~\eqref{eq:SM_pq_recursion} with initial condition \(p_0^{(0)}=1\) and \(p_q^{(0)}=0\) for \(q\neq 0\).

Next, we compute the conditional occupation of the uniform orbital
\begin{align}
    \nu_q
    =
    \Tr(\rho_{A,q}n_{A,0}),
    \qquad
    \rho_{A,q} = \frac{P_q \rho_A P_q}{p_q},
    \qquad
    n_{A,0}=d_{A,0}^\dagger d_{A,0},
    \qquad
    d_{A,0}
    =
    \frac{1}{\sqrt{N_A}}
    \sum_{i\in A}c_i .
\end{align}
In the \(f_\alpha\) basis, the uniform orbital can be written as 
\begin{align}
    d_{A,0}
    =
    \sum_{\alpha=1}^{N_A}
    s_\alpha f_\alpha,
    \qquad
    s_\alpha
    =
    \frac{1}{\sqrt{N_A}}
    \sum_{i\in A}
    u_{\alpha i}^{*}.
\end{align}
Since \(\rho_{A,q}\) is diagonal in the occupation basis of the \(f_\alpha\) modes, we obtain 
\begin{align}\label{eq:Tr[rhoff]}
\Tr[\rho_{A,q}f_\alpha^\dag f_\beta] = \delta_{\alpha,\beta} \Pr(n_\alpha=1|Q_A=q), 
\end{align}
where $\Pr(n_\alpha=1\mid Q_A=q)$ is the probability that the $\alpha$-th eigenmode is occupied, conditioned on finding $q$ particles in subsystem $A$. 
Using Eq.~\eqref{eq:Tr[rhoff]}, $\nu_q$ can be written as 
\begin{align}
    \nu_q
    =
    \sum_{\alpha=1}^{N_A}
    |s_\alpha|^2
    \Pr(n_\alpha=1\,|\,Q_A=q).
    \label{eq:SM_nuq_conditional}
\end{align}

The conditional probabilities in Eq.~\eqref{eq:SM_nuq_conditional} are computed as follows. For each \(\alpha\), we define
\begin{align}
    F^{(\alpha)}(z)
    =
    \prod_{\beta\neq\alpha}
    \left[
    (1-\lambda_\beta)+\lambda_\beta z
    \right]
    =
    \sum_{m=0}^{N_A-1}
    p_m^{(\alpha)} z^m .
\end{align}
Then the joint probability of \(n_\alpha=1\) and \(Q_A=q\) is
\begin{align}
    \Pr(n_\alpha=1,Q_A=q)
    =
    \lambda_\alpha p_{q-1}^{(\alpha)}.
\end{align}
Therefore,
\begin{align}
    \nu_q
    =
    \frac{1}{p_q}
    \sum_{\alpha=1}^{N_A}
    |s_\alpha|^2
    \lambda_\alpha
    p_{q-1}^{(\alpha)} ,
    \label{eq:SM_nuq_final}
\end{align}
where \(p_{-1}^{(\alpha)}=0\). In the numerical calculation, the coefficients \(p_m^{(\alpha)}\) are obtained from the same recurrence as Eq.~\eqref{eq:SM_pq_recursion}, with the mode \(\alpha\) omitted.

\ifdefined\includedinsupplement
\let\finishsupplement 
\else
\let\finishsupplement\relax
\fi
\finishsupplement

\bibliographystyle{./apsrev4-2-title}
\bibliography{ref}

\end{document}